\documentstyle[preprint,revtex]{aps}

\begin{document}
\preprint{IFUSP/P-1002}
\preprint{hep-th/9208049}
\draft

\begin{title}
Pseudoclassical Model of Spinning Particle \\ with Anomalous Magnetic Momentum
\end{title}

\author{D.M.Gitman and A.V.Saa}
\begin{instit}
Instituto de F\'{\i}sica\\
Departamento de F\'{\i}sica Matem\'atica\\
Universidade de S\~ao Paulo, Caixa Postal 20516 \\
01498-970 S\~ao Paulo, S.P.\\
Brazil
\end{instit}

\begin{abstract}
A generalization of the pseudoclassical
action of a spinning particle in  the presence of an
anomalous magnetic momentum is given. The action is written in
reparametrization and supergauge invariant form. The Dirac quantization, based
on the Hamiltonian analyses of the model, leads to the
Dirac-Pauli equation for a particle with an anomalous magnetic
momentum in an external electromagnetic field.
Due to the structure of first-class constraints in that case,
the Dirac quantization demands for consistency to take into account
an operators ordering problem.
\end{abstract}


In recent years numerous classical models of relativistic
particles and superparticles  were discussed
intensively.
First, the interest to such models was initiated by the
close relation
with the string theory, but now it is also clear that the problem itself
has an important meaning for the
 deeper understanding of the structure of quantum
theory.
One of the basic, in the above mentioned set of classical models, is the
pseudoclassical model of Fermi particle with  spin $1/2$, proposed first
in the works \cite{ber,casal}, investigated and quantized in many
works, see for example
 \cite{ber,casal,brink,vechia,hen,sund,mann,gittyu,gitfrad}.
The model can be formulated in gauge
invariant (reparametrization and supersymmetric) form.
The action of the model in an external electromagnetic
field has the form \cite{gitfrad}:
\begin{equation}
 S = \int_0^1 \left[
 -\frac{\dot{x}^2}{2e}  - e\frac{m^2}{2} - g\dot{x}^\alpha A_\alpha
+igeF_{\alpha\beta}\psi^\alpha\psi^\beta
 +i\left(\frac{\dot{x}_\alpha\psi^\alpha}{e} -
m\psi^5
\right)\chi -i\psi_n\dot{\psi}^n
\right]d\,\tau\, ,
\label{lag}
\end{equation}
where $x^\alpha$, $e$ are even and $\psi^n$, $\chi$ are odd variables dependent
on $\tau$, which plays the role of the time in this theory,
$A_\alpha(x)$ is an external electromagnetic field potential,
 $F_{\alpha\beta}(x)$
is the Maxwell strength tensor, and g the electrical charge. Greek indices
run over $\overline{0,3}$ and Latin indices run over $\overline{0,3},5$.
The metric tensors:
$\eta_{\alpha\beta}= {\rm diag}(1,-1,-1,-1)$ and
$\eta_{mn}= {\rm diag}(1,-1,-1,-1,-1)$.
 There are two gauge
transformations in the theory with the action (\ref{lag}), reparametrizations,
\begin{equation}
\delta x = \dot{x}\xi \,\,,\,\,\,\,\,\,
\delta e = \frac{d}{dt}(e\xi) \,\,,\,\,\,\,\,\,
\delta \psi^n = \dot{\psi}^n\xi \, \,,\,\,\,\,\,\,
\delta \chi = \frac{d}{dt}(\chi\xi)\, ,
\label{re}
\end{equation}
and supertransformations,
\begin{eqnarray}
&&\delta x = i\psi\epsilon \,\,,\,\,\,\,\,\,
\delta e = i\chi \epsilon \,\,,\,\,\,\,\,\,
\delta \chi = \dot{\epsilon} \,\,,\,\,\,\,\,\,
\delta\psi^\alpha = \frac{1}{2e}(\dot{x}^\alpha
+ i\chi\psi^\alpha)\epsilon \,\, ,  \nonumber \\
&&\delta\psi^5 = \left[\frac{m}{2}
-\frac{i}{me}\psi^5\left(
\dot{\psi}^5
-\frac{m}{2}\chi
\right)
\right]\epsilon\,\,,
\label{su}
\end{eqnarray}
where $\xi$ are even and $\epsilon$ odd $\tau$-dependent parameters.
The spinning degrees
of freedom in such a model are described by Grassmannian
 variables, that's why the model is called pseudoclassical.
The quantization of the model in different ways  leads to the
quantum mechanics of the Dirac particle, is very instructive and
creates many useful analogies with problems of quantization of gauge
field theories.

In this letter we present a generalization of the
model  when  an anomalous magnetic momentum of the particle is present.
The relativistic quantum theory of a spinning particle, which has both the
``normal'' magnetic momentum $g/2m$ and an ``anomalous'' magnetic
momentum $\mu$, was formulated by Pauli \cite{pauli}. In this case he
generalized the Dirac equation to the following form:
\begin{equation}
 \left( \hat{{\cal P}}_\nu \gamma^\nu - m -
\frac{\mu}{2}\sigma^{\alpha\beta}
F_{\alpha\beta} \right)\Psi(x)= 0,
\label{gendirac}
\end{equation}
where
 $\hat{{\cal P}}_\nu = i \partial_\nu - g A_\nu(x)$,
$\sigma^{\alpha\beta}=
\frac{i}{2}[\gamma^\alpha,\gamma^\beta]_-$,
$[\gamma^\alpha,\gamma^\beta]_+ = 2\eta^{\alpha\beta}$, and the notations
$[A,B]_\pm = AB \pm BA$ are used.

\noindent
Our aim is to write an analog of the action (\ref{lag}), whose quantization
gives the Dirac-Pauli theory.

We propose the following pseudoclassical
action for a spinning particle with an anomalous momentum:
\begin{eqnarray}
 S &=& \int_0^1 \left[
 -\frac{\dot{x}^2}{2e}  - e\frac{M^2}{2} - \dot{x}^\alpha\left(gA_\alpha
+4i\mu\psi^5 F_{\alpha\beta}\psi^\beta \right)
+igeF_{\alpha\beta}\psi^\alpha\psi^\beta
  \right.\nonumber \\
&&\left. +i\left(\frac{\dot{x}_\alpha\psi^\alpha}{e} -
M^*\psi^5
\right)\chi -i\psi_n\dot{\psi}^n
\right]d\,\tau\, ,
\label{lagr}
\end{eqnarray}
where $M=m-2i\mu F_{\alpha\beta}\psi^\alpha\psi^\beta$, and
$M^*=m+2i\mu F_{\alpha\beta}\psi^\alpha\psi^\beta$.

\noindent
One can check that there are also two types of gauge transformations, under
which the actions is invariant.
 The first one, which is the reparametrization, has the
same form as (\ref{re}). The second one is a supertransformation,
\begin{eqnarray}
&&\delta x = i\psi\epsilon \,\,,\,\,\,\,\,\,
\delta e = i\chi \epsilon \,\,,\,\,\,\,\,\,
\delta \chi = \dot{\epsilon} \,\,,\,\,\,\,\,\,
\delta\psi^\alpha = \frac{1}{2e}(\dot{x}^\alpha
+ i\chi\psi^\alpha)\epsilon \,\, ,  \nonumber \\
&&\delta\psi^5 = \left[\frac{M}{2}
-\frac{i}{me}\psi^5\left(
\dot{\psi}^5 - 2\mu F_{\alpha\beta}\dot{x}^\alpha\psi^\beta
-\frac{M^*}{2}\chi
\right)
\right]\epsilon\,\,,
\label{sup}
\end{eqnarray}
 The form of
the transformation (\ref{sup}) depends on the
external field and the anomalous momentum and generalizes the
transformation (\ref{su}).

 Going over to the
Hamiltonian formalism, we introduce  canonical momenta:
\begin{eqnarray}
&&p_\alpha = \frac{\partial L}{\partial\dot{x}^\alpha} = -\frac{1}{e}\left(
\dot{x}_\alpha - i\psi_\alpha\chi\right) - gA_\alpha
- 4i\mu\psi^5 F_{\alpha\beta}\psi^\beta, \nonumber \\
&&P_e = \frac{\partial L}{\partial\dot{e}} = 0, \,\,\,\,\,\,\,\,
P_\chi = \frac{\partial_r L}{\partial\dot{\chi}} = 0, \,\,\,\,\,\,\,\,
P_n =  \frac{\partial_r L}{\partial\dot{\psi}^n} = -i\psi_n\,\,\,.
\label{momenta}
\end{eqnarray}
It follows from the
equation  (\ref{momenta}) that there exist primary constraints
 $\Phi_a^{(1)}=0$,
\begin{equation}
\Phi_a^{(1)}=\left\{
\begin{array}{l}
\Phi_1^{(1)}= P_\chi\,\,\,, \\
\Phi_2^{(1)}= P_e\,\,\,, \\
\Phi_{3n}^{(1)}= P_n +i\psi_n\,\,\,.
\end{array}
\right.
\end{equation}
We construct the  Hamiltonian $H^{(1)}$, according to the standard procedure
(we use the notations of the book \cite{quan}),
\begin{equation}
H^{(1)}= H + \lambda_a \Phi_a^{(1)},\,\,\, {\rm where}\,\,\,\,
H = \left.\left(
\frac{\partial_rL}{\partial\dot{q}}\dot{q}-L\right)
\right|_{\frac{\partial_rL}{\partial\dot{q}}=P},
\,\,\,\,\,\, q = (x,e,\chi,\psi^n)\,\,\,,
\end{equation}
and get for $H$:
\begin{equation}
H = -\frac{e}{2} \left(
{\cal P}^2 - 8i\mu \psi^5F_{\alpha\beta}{\cal P}^\alpha\psi^\beta  +
2igF_{\alpha\beta}\psi^\alpha\psi^\beta  - M^2
\right) -i\left(  {\cal P}_\alpha\psi^\alpha -M\psi^5 \right)\chi\, ,
\end{equation}
where ${\cal P}_\alpha = -p_\alpha - gA_\alpha$.

\noindent
{}From the
 conditions of the conservation of the primary constraints $\Phi_{1,2}^{(1)}$
in time $\tau$ ,
$\dot{\Phi}^{(1)}_{1,2} = \left\{{\Phi}^{(1)}_{1,2},H^{(1)} \right\} = 0$,
we find the secondary constraints $\Phi_{1,2}^{(2)} = 0$,
\begin{eqnarray}
\label{first}
\Phi_1^{(2)}&=& {\cal P}_\alpha\psi^\alpha -M\psi^5 =0,  \\
\Phi_2^{(2)}&=&
{\cal P}^2 - 8i\mu \psi^5F_{\alpha\beta}{\cal P}^\alpha\psi^\beta
+2igF_{\alpha\beta}\psi^\alpha\psi^\beta -  M^2=0,
\label{second}
\end{eqnarray}
and  the same conditions for the constraints $\Phi^{(1)}_{3n}$ give
equations for the determination of $\lambda_{3n}$.
Thus, the Hamiltonian $H$ appears to be
 proportional to constraints, as one can
expect in the case of a re\-pa\-ra\-me\-tri\-za\-tion invariant theory,
\begin{equation}
H =  i\chi\Phi^{(2)}_1 -\frac{e}{2} \Phi^{(2)}_2 .
\end{equation}
No more secondary constraints  arise from the Dirac procedure, and the
Lagrange's multipliers $\lambda_{1}$ and $\lambda_{2}$ remain undetermined,
in perfect correspondence with the fact that the number of gauge
transformations parameters equals two for the theory in question \cite{quan}.
Now we go over from the initial set of constraints
$\left(\Phi^{(1)},\Phi^{(2)}\right)$ to the equivalent one
$\left(\Phi^{(1)},T\right)$, where
\begin{equation}
 T = \Phi^{(2)} +
\frac{i}{2} \frac{\partial_r\Phi}{\partial\psi^n}^{(2)}\Phi^{(1)}_{3n}\,\,,
\label{dt}
\end{equation}
because of the latter  can be explicitly divided in
a set of first-class constraints, which is
$\left(\Phi^{(1)}_{1,2},T\right)$ and second-class ones, which is
$\Phi^{(1)}_{3n}$.
In our case  we perform only a partial gauge fixing, by imposing
 the supplementary gauge conditions
$\Phi^{\rm G}_{1,2}=0$ to the primary first-class constraints
$\Phi^{(1)}_{1,2}\,\,$,
\begin{equation}
\Phi^{\rm G}_1 = \chi=0, \,\,\,\,\,\,
\Phi^{\rm G}_2 = e = 1/m.
\label{gauge}
\end{equation}
One can check that the conditions of the conservation in time of the
supplementary constraints
(\ref{gauge}) give equations for determination of the
multipliers $\lambda_1$ and $\lambda_2$.
Thus, on this stage we reduced our Hamiltonian theory to one
with the first-class constraints $T$ and second-class ones
$\varphi = \left(\Phi^{(1)},\Phi^{\rm G} \right)$.

For the quantization we will use the so called Dirac method for
 systems with first-class constraints \cite{yesh}, which, being
generalized to the presence of second-class constraints, can be
formulated as follow: the commutation relations between operators are
calculated according to the Dirac brackets with respect to the
second-class  constraints only; second-class constraints operators
equal zero; first-class constraints as operators are not zero, but,
are considered in sense of restrictions on state vectors.
All the operators equations have to be realized in some Hilbert space.

In our case, the sub-set of second-class constraints
$\left(\Phi^{(1)}_{1,2},\Phi^{\rm G}\right)$ has a special form \cite{quan},
so that one can use it for eliminating of the variables $e,P_e,\chi,P_\chi$,
 from the consideration, then,
for the rest of the variables $x,p,\psi^n$,
the Dirac brackets with respect to the constraints $\varphi$
reduce to ones with respect to the constraints $\Phi^{(1)}_{3n}$ only and
can be easy calculated,
\begin{equation}
 \left\{x^\alpha,p_\beta \right\}_{D(\Phi^{(1)}_{3n})} =
 \delta^\alpha_\beta \,,\,\,\,\,\,\,\,
 \left\{\psi^n,\psi^m \right\}_{D(\Phi^{(1)}_{3n})} =
\frac{i}{2}\eta^{nm}\,,
\end{equation}
while  others Dirac brackets vanish.
Thus, the commutation relations for the operators
$\hat{x},\hat{p},\hat{\psi}^n$, which correspond to the variables
$x,p,\psi^n$ respectively, are
\begin{equation}
 \left[\hat{x}^\alpha,\hat{p}_\beta \right]_-
= i\left\{x^\alpha,p_\beta \right\}_{D(\Phi^{(1)}_{3n})} =
\delta^\alpha_\beta, \,\,\,\,\,
 \left[\hat{\psi}^m,\hat{\psi}^n \right]_+
= i\left\{\psi^m,\psi^n \right\}_{D(\Phi^{(1)}_{3n})}=
-\frac{1}{2}\eta^{mn}.
\label{comm}
\end{equation}
Besides, the operator equations hold:
\begin{equation}
\hat{\Phi}^{(1)}_{3n}=  \hat{P}_n + i \hat{\psi}_n =0 .
\label{oe}
\end{equation}
The commutation relations (\ref{comm}) and the equations
 (\ref{oe}) can be realized
in a space of the four columns $\Psi(x)$ dependent on $x^\alpha$
as: $\hat{x}^\alpha$ are
operators of multiplication, $\hat{p}_\alpha = -i\partial_\alpha$,
$\hat{\psi}^\alpha = \frac{i}{2}\gamma^5\gamma^\alpha$, and
$\hat{\psi}^5 = \frac{i}{2}\gamma^5$,
where $\gamma^n$ are the $\gamma$-matrices
$(\gamma^\alpha,\gamma^5)$,
$\left[\gamma^m,\gamma^n \right]_{+}=2\eta^{mn}$.
The first-class constraints $\hat{T}$ as operators have to annihilate
physical vectors; in virtue of (\ref{oe}), (\ref{dt}) these conditions
 reduce to the equations:
\begin{equation}
\hat{\Phi}^{(2)}_{1,2}\Psi(x)=0,
\label{t}
\end{equation}
where $\hat{\Phi}^{(2)}_{1,2}$ are operators, which correspond to the
constraints (\ref{first}), (\ref{second}). There is no ambiguity in
construction of the operator $\hat{\Phi}^{(2)}_1$, according to the classical
function $\Phi^{(2)}_1$ from (\ref{first}). Thus, taking into account
the realizations of the commutation relations (\ref{comm}), one easily can see
that the first equation  (\ref{t})
reproduces the Dirac-Pauli equation  (\ref{gendirac}).
As to the construction of the operator $\hat{\Phi}^{(2)}_2$, according to the
classical function ${\Phi}^{(2)}_2$ from (\ref{second}), we meet here an
ordering problem since the constraint ${\Phi}^{(2)}_2$ contains terms
with products of the momenta and functions of the coordinates, namely terms
of the form $p_\alpha A^\alpha$,
$p_\alpha F^{\alpha\beta}$.
For such  terms we choose the symmetrized form of the
corresponding operators,
\begin{equation}
p_\alpha A^\alpha \rightarrow \frac{1}{2}
\left[\hat{p}_\alpha,A^\alpha(\hat{x}) \right]_{+} \, , \,\,\,\,\,\,\,\,
p_\alpha F^{\alpha\beta} \rightarrow \frac{1}{2}
\left[\hat{p}_\alpha,F^{\alpha\beta}(\hat{x}) \right]_{+} \, ,
\label{real}
\end{equation}
which, in particular, provides the hermiticity of the operator
$\hat{\Phi}^{(2)}_2$. But the main reason is, the correspondence rule
(\ref{real}) provides the consistency of the two equations (\ref{t}). Indeed,
in this case we have
\begin{equation}
\hat{\Phi}^{(2)}_2 = \left(\hat{\Phi}^{(2)}_1 \right)^2,
\label{sq}
\end{equation}
and the second equation  (\ref{t}) appears to be merely the consequence of the
first equation  (\ref{t}), i.e. of the Dirac-Pauli equation  (\ref{gendirac}).
To verify the validity of (\ref{sq}), one needs only to take into
 account that the operator, which corresponds to the term
$8 i\mu \psi^5F_{\alpha\beta}{\cal P}^\alpha\psi^\beta$
in the constraint
${\Phi}^{(2)}_2$ (\ref{second}), in virtue of the structure of the
$\gamma$-matrices, can be written in the form:
\begin{eqnarray}
&& 8 i\mu \psi^5F_{\alpha\beta}{\cal P}^\alpha\psi^\beta \rightarrow
i\mu\left[{F}_{\alpha\beta}(\hat{x}),\hat{{\cal P}}^\alpha
\right]_{+}\gamma^\beta = \nonumber \\
&& 2i\mu F_{\alpha\beta}(\hat{x})\hat{\cal P}^\alpha\gamma^\beta +
\mu\partial^\alpha F_{\alpha\beta}(\hat{x})\gamma^\beta =
\left[\hat{\cal P}_\alpha\gamma^\alpha, \frac{\mu}{2}
\sigma^{\alpha\beta}F_{\alpha\beta}(\hat{x}) \right]_{-}
\end{eqnarray}
To complete the quantization, one has to present an inner product in the
space of realization of commutation relations. The general method of its
construction, in the frame of the Dirac method  we used,
is unfortunately still unknown. Nevertheless, in this concrete case,
the space of physical vectors, obeying the condition
(\ref{t}), can be transformed into a Hilbert space, if one
takes for the inner product ordinary  scalar product of solutions of
the Dirac equation, which does not depend on $x^0$, in spite of the
integration is fulfilled in it over $x^i$ only.
It is not
difficult to verify that the introduced operators, obey of natural properties
of hermiticity, which are known from the Dirac relativistic mechanics. In
particular, the operator $\hat{p}_0$, which has to be considered on the same
 foot with $\hat{p}_i$, is also hermitian on the solutions of the
Dirac-Pauli equation (\ref{gendirac}), in virtue of the above mentioned
independence of the scalar product on  $x^0$.

The authors would like to thank Professor Josif Frenkel for discussions and
one of them (AVS) thanks FAPESP(Brazil) for support.

\end{document}